\documentclass[twocolumn,prl,showpacs]{revtex4} 
\usepackage{graphicx,amssymb,amsmath,bm}

\begin{document}

\title{Resonant inelastic x-ray scattering on single-magnons at oxygen $K$-edges}

\author {Beom Hyun Kim}
\author {Jeroen van den Brink}
\affiliation{Institute for Theoretical Solid State State Physics, IFW Dresden,
Helmholtzstrasse 20, 01069 Dresden, Germany}
\date{\today}

\begin{abstract}
The recent discovery that resonant inelastic x-ray scattering can probe 
single-magnon (SM) dispersions in transition metal (TM) oxides 
when the x-ray energy is tuned to the TM $L$-edge, 
has put this technique on a par with inelastic neutron scattering. 
It is generally presumed that selection rules forbid SM-scattering 
at oxygen (O) $K$-edges. 
However, based on a symmetry analysis and exact diagonalization study, 
we show that SM-scattering at O $K$-edges becomes allowed
when ($i$) spin-orbit coupling is present in the TM $d$-shell and 
($ii$) inversion symmetry at the O-site is broken. 
For cuprates the resulting SM-amplitude is very weak but in iridates both prerequisites can amply be fulfilled.
\end{abstract}


\maketitle

{\it Introduction ---}
The theoretical prediction~\cite{Ament2009a} and 
subsequent experimental observation~\cite{Braicovich2009a,Braicovich2009b} 
that the dispersion of magnetic excitations, 
in particular of elementary single-magnons, can be measured directly 
by resonant inelastic x-ray scattering (RIXS) has fundamentally changed 
the field of inelastic magnetic scattering~\cite{Ament2011}. 
It has for instance lead to the discovery of distinct paramagnons 
in a large family of high-temperature cuprate superconductors
\cite{Braicovich2009b,Ament2011,Tacon2011,Tacon2013,Dean2013a,Dean2013b}, 
and established the presence of strongly dispersive magnetic modes 
in a number of iridium oxides~\cite{Ament2011b,Kim2012a,Kim2012b,Gretarsson2013}.

All these efforts to extract the dispersion of elementary magnetic excitations 
from RIXS have focussed on the transition metal (TM) $L$-edge, 
where single spin-flip scattering is directly allowed
\cite{Ament2009a,Braicovich2009a,Braicovich2009b,Ament2011,Tacon2011,Tacon2013,Dean2013a,Dean2013b,Ament2011b,Kim2012a,Kim2012b,Gretarsson2013,Haverkort2009,Veenendaal2011,Igarashi2013a,Igarashi2013b}. 
The microscopic origin of this type of magnetic scattering is rather straightforward. 
When energy in RIXS the incoming photon is tuned to a TM $L$-edge, 
an electron with spin $\sigma$ is excited out of the atomic $p$-shell, 
deep inside the atomic core, into the TM $d$-shell. 
The core-hole that is created in this event can now flip its spin 
due its very large spin-orbit interaction. 
Subsequently a valence $d$ electron with spin $-\sigma$ can fill 
the core-hole and an outgoing x-ray is emitted. 
The net result of this RIXS process is a spin-flip transition 
$\sigma \rightarrow -\sigma$ in the TM $d$-shell, 
which is the same as the net result of inelastic neutron scattering 
involving this $d$-shell electron. 

When a $K$-edge is used instead of a $L$-edge, 
this direct spin-flip process is no longer allowed. 
This is because in $K$-edge RIXS an electron is excited 
from a core-shell with $s$-symmetry, for which spin-orbit coupling is simply absent. 
It thus appears that for RIXS at $K$-edges direct spin-flip scattering is forbidden 
and only higher order magnetic scattering processes 
-- starting at double spin-flip (e.g. bi-magnon) scattering -- 
are allowed~\cite{Brink2007,Nagao2007,Forte2008a}. 
Indeed bi-magnon scattering is observed both at TM~\cite{Hill2008,Ellis2010} and 
oxygen (O) $K$-edges~\cite{Bisogni2012a,Bisogni2012b}. 
This apparent absence of single-magnon spin-flip scattering is unfortunate 
because the O $K$-edge is in a soft x-ray regime 
where the RIXS resolution is particularly high 
(even if in this situation only a limited part of momentum space is accessible) 
and oxygen anions are obviously ubiquitous in magnetic materials.

However, in this Letter we show that single-magnon (SM) scattering at O $K$-edges 
is allowed when spin-orbit coupling is present 
in the TM $d$-shell {\it and} inversion symmetry at the O-site is broken. 
Our symmetry analysis and exact diagonalization study shows that 
the resulting SM-scattering is allowed for small momentum losses 
and forbidden for momenta close to 
the magnetic ordering vector, a situation opposite to magnetic RIXS at the TM $L$-edge. 
As in iridates the spin-orbit interaction in the $5d$ shell is large and 
the inversion symmetry at O-sites is often distinctly broken, as e.g. 
in the perovskites Sr$_2$IrO$_4$, Sr$_3$Ir$_2$O$_7$, Na$_2$IrO$_3$ and 
the post-perovskite CaIrO$_3$, 
we predict pronounced SM-scattering at the O $K$-edge of for instance 
these magnetic oxides.

{\it RIXS cross section ---}
When an incident (outgoing) x-ray with $\mathbf{\epsilon}$ 
($\mathbf{\epsilon}^{\prime}$) polarization
has the frequency $\omega_{\mathbf{k}}$ ($\omega_{\mathbf{k}^{\prime}}$)
and wave vector $\mathbf{k}$ ($\mathbf{k}^{\prime}$),
the RIXS intensity is described by the Kramers-Heisenberg relation~\cite{Ament2011}
\begin{multline}
I\left(\omega,\mathbf{q},\mathbf{\epsilon},\mathbf{\epsilon}^{\prime}
  \right)=\sum_f |\mathcal{F}_{fg}(\omega,\mathbf{q},
 \mathbf{\epsilon},\mathbf{\epsilon}^{\prime})|^2 \\
  \times \delta\left(E_f+\hbar\omega_{\mathbf{k}}-E_g-
  \hbar\omega_{\mathbf{k}^{\prime}}\right),
\end{multline}
where, $\omega=\omega_{\mathbf{k}}-\omega_{\mathbf{k}^{\prime}}$
and $\mathbf{q}=\mathbf{k}-\mathbf{k}^{\prime}$.
In the dipole and fast collision approximation, 
the scattering amplitude $\mathcal{F}_{fg}$ is reduced to
$
\mathcal{F}_{fg}=\frac{1}{i\Gamma} \langle 
f|R(\mathbf{\epsilon}^{\prime},\mathbf{\epsilon},\mathbf{q}) | g \rangle,
$
where $R$ is the effective RIXS scattering operator, 
which is defined by the product of two dipole operators, and 
$\Gamma$ is the core-hole broadening \cite{Ament2011}. 
When $\psi_{\nu}$ and $\psi_s$ represent valence and core states, 
the scattering operator is given by
\begin{equation}
\label{R_eq}
R(\mathbf{\epsilon}^{\prime},\mathbf{\epsilon},\mathbf{q})=
\sum_{i\nu\nu^{\prime}} e^{i\mathbf{q}\cdot\mathbf{r}_i}
 T_{\nu^{\prime}\nu}(\mathbf{\epsilon}^{\prime},\mathbf{\epsilon})
 c_{i\nu^{\prime}}c_{i\nu}^{\dagger},
\end{equation}
where $T_{\nu^{\prime}\nu}(\mathbf{\epsilon}^{\prime},\mathbf{\epsilon})=
 \sum_{s} \langle \psi_{\nu^{\prime}} | \mathbf{\epsilon}^{\prime}\cdot \mathbf{r} 
 |\psi_s\rangle
  \langle \psi_s | \mathbf{\epsilon}\cdot \mathbf{r} |\psi_{\nu}\rangle$.
In case of the O $K$-edge RIXS, 
$T_{\nu^{\prime}\nu}(\mathbf{\epsilon}^{\prime},\mathbf{\epsilon})$  equals
$\frac{1}{3}\epsilon_{m^{\prime}} \epsilon_{m} \delta_{\sigma^{\prime},\sigma}
 \langle r \rangle^2$ where $\nu$ has $m$ ($x$, $y$, or $z$) orbital and 
 $\sigma$ spin characters. The most important feature of the O $K$-edge RIXS process 
is that in its intermediate state, besides an oxygen core-hole, 
also an extra valence electron is present the system that 
hybridizes and interacts with the TM $d$-electrons.

\begin{figure}[t]
\centering
\includegraphics[width=.7\columnwidth]{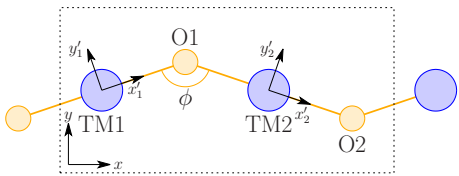}
\caption
{ (Color online)  Schematic diagram of corner-sharing TM-O system, 
which for bond-angle $\phi$ not equal to 180$^{\circ}$ lacks inversion symmetry 
at the oxygen site. The $z$-direction is out of plane.   
}
\label{fig1}
\end{figure}

{\it Symmetry analysis ---}
To investigate the magnetic response of the O $K$-edge RIXS, 
we start by considering a simple TM1-O1-TM2-O2  arrangement of transition metal and 
oxygen ions, where the TM-O-TM bond angle is $\phi$, see Fig.~\ref{fig1}. 
To be specific, we concentrate on iridates (cuprates) in the following symmetry analysis, 
in which the Ir$^{4_+}$ (Cu$^{2+}$) ions are in a $5d^5$ ($3d^9$) configuration. 
Due to the strong spin-orbit coupling, Ir$^{4_+}$ ions in iridate
form effective $J_{eff}=1/2$ moments~\cite{BJKim2008,BJKim2009} so that 
the single-site ground-state multiplet is a Kramers doublet, which we denote by $D$. 
Also in the case of Cu$^{2+}$ in cuprates, a Kramers doublet ($S=1/2$) is the ground state 
as the tetragonal crystal field splits up the $e_g$ orbital degeneracy. 
When one considers two of these moments, the magnetic interactions between them 
cause a splitting into a total singlet ($S$) and a total triplet ($T$) state. 
The four states are
$|S\rangle=\frac{1}{2}\left(|D_{\uparrow}D_{\downarrow};00\rangle-
     |D_{\downarrow}D_{\uparrow};00\rangle \right)$,
$|T_1\rangle=|D_{\uparrow}D_{\uparrow};00\rangle$, 
$|T_0\rangle=\frac{1}{2}\left(|D_{\uparrow}D_{\downarrow};00\rangle+
     |D_{\downarrow}D_{\uparrow};00\rangle \right)$, and
$|T_{-1}\rangle=|D_{\downarrow}D_{\downarrow};00\rangle$,
where $|00\rangle$ denotes the {\it absence of holes} on the oxygen sites, 
in the O $p$-orbitals. Because of the hopping $t_{pd}$ between TM and O, 
such charge-transfer states of course mix in. We denote such a charge-transfer state 
with a hole on oxygen as $|D_{\sigma}A;\xi0\rangle$, 
where $A$ represents the Ir $5d^6$ (Cu $3d^{10}$) state, 
in which all doublet states are occupied, and $\xi$ refers to the $p$-hole state of O.

We now introduce two different inversion symmetry operators 
$I_{\textrm{O}}$ and $I_{TM}$ that represent inversion 
with respect to O and TM centers, respectively. 
When $I_{\text{O}}$ and $I_{TM}$ are applied, local coordinates transform as 
%
$
I_{\textrm{O}} :  ~(x_1^{\prime},y_1^{\prime})  \rightarrow (-x_2^{\prime},-y_2^{\prime}),   (x_{1},y_{1}) \rightarrow (-x_{1},-y_{1}) 
$
and
$
I_{TM} :  ~(x_1^{\prime},y_1^{\prime}) \rightarrow  (-x_1^{\prime},-y_1^{\prime}),  (x_{1},y_{1}) \rightarrow (-x_{2},-y_{2}).
$
%
We note that $I_{TM}$ is {\it always} a symmetry of the system, 
but $I_{\textrm{O}}$ is a symmetry operator {\it only for $\phi=180^{\circ}$}, 
i.e. undistorted TM-O-TM bonds. Because of $I_{TM}$ symmetry, 
the hopping integrals between singlet/triplet and charge-transfer states
obey the relations
$
\langle S|H_t|D_{\sigma}A;\xi0\rangle =-\langle S|H_t|D_{\sigma}A;0\xi\rangle
$
and 
$
\langle T_m|H_t|D_{\sigma}A;\xi0\rangle =-\langle T_m|H_t|D_{\sigma}A;0\xi\rangle,
$
where $H_t$ is the TM-O hopping part of the Hamiltonian, so that the singlet ground state $|g\rangle$ and triplet excites states $|m\rangle$ become 
%
%
\begin{align}
|g\rangle &= c |S\rangle + \sum_{\sigma\xi} c_{\sigma\xi}  \big( |D_{\sigma}A;\xi0\rangle -|D_{\sigma}A;0\xi\rangle \big)+\cdots, \nonumber \\
|m\rangle &= c_m |T_m\rangle + \sum_{\sigma\xi} c_{m\sigma\xi} \big( |D_{\sigma}A;\xi0\rangle -|D_{\sigma}A;0\xi\rangle \big)+\cdots, \nonumber 
\end{align}
which directly follows from the symmetry properties of the states involved given explicitly in Table I.

\begin{table}[t]
\centering
\caption
{Transformation properties of singlet ($|S\rangle$), triplet ($|T\rangle$), 
and  charge transfer ($|D_{\sigma}A;\xi0\rangle$)  states 
under two types of inversion, $I_{\textrm{O}}$ and $I_{TM}$. 
Note that $I_{\textrm{O}}$ is only a symmetry 
when the bond angle $\phi=180^{\circ}$.
} \label{table1}
\begin{ruledtabular}
\begin{tabular}{l | r r r}
                                     &  $|S\rangle$ & $|T\rangle$ & 
    $|D_{\sigma}A;\xi0\rangle$ \\
\hline
 $I_{\textrm{O}} (\phi=180^{\circ})$ &  $-|S\rangle$ & $|T\rangle$ & 
    $-|AD_{\sigma};\xi0\rangle$ \\
 $I_M$                               &  $|S\rangle$ & $|T\rangle$ & 
    $-|DA_{\sigma};0\xi\rangle$ \\
\end{tabular}
\end{ruledtabular}
\end{table}

Next we require the symmetry properties of the effective RIXS scattering operator 
$R(\epsilon^\prime,\epsilon,{\bf q})$ in Eq.~(\ref{R_eq}) under inversion. 
These explicitly depend on the momentum $\bf q$ that is transferred 
by a x-ray photon to the system: only for momenta corresponding to 
high symmetry points in the Brillouin Zone (BZ), a symmetry analysis is viable, 
in the present case in particular for $q=0$ and $q=\pi$. 
For these momenta the transformation properties of $R$ 
under the two types of inversion, $I_{\textrm{O}}$ and $I_{TM}$, are summarized in Table II, 
considering both RIXS at the O $K$-edge and the Ir $L_3$-edge. 

\begin{table}[b]
\centering
\caption
{Symmetry properties of RIXS scattering operator $R$ at the O $K$-edge and 
Ir $L_3$-edge with respect to two types of inversions, 
$I_{\textrm{O}}$ and $I_{TM}$.
} \label{table2}
\begin{ruledtabular}
\begin{tabular}{l | c c | c c}
\        & \multicolumn{2}{c|}{Ir $L$-edge} & \multicolumn{2}{c}{O $K$-edge} \\
        &  $q=0$  &  $q=\pi$  &  $q=0$  &  $q=\pi$ \\
\hline
 $I_{\textrm{O}} (\phi=180^{\circ})$ &  $R$ & $-R$ &  $R$ & $\quad\! R$ \\
 $I_{TM}$ &  $R$ & $\quad\! R$ &  $R$ & $-R$ \\
\end{tabular}
\end{ruledtabular}
\end{table}

Having determined these transformation properties, 
it can immediately verified that the RIXS scattering amplitude 
$\mathcal{F}_{mg}^{O K}$ at the O $K$-edge between ground state $|g\rangle$ and 
magnetically excited state $|m\rangle$ obeys 
the selection rule $\mathcal{F}_{mg}^{O K}\left(q=\pi\right)=0$ for all $m$, 
because  $I_{TM}^{\dagger}R(q=\pi)I_{TM}=-R$. 
For $q=0$ no such selection rule is dictated by $I_{TM}$. 
However, in the situation that the bond-angle is $180^{\circ}$ and 
$I_{\textrm{O}}$ is a symmetry of the system, 
the relations $\langle S|H_t|D_{\sigma}A;\xi0\rangle =\langle S|H_t|AD_{\sigma};\xi0\rangle $ 
and $\langle T_m|H_t|D_{\sigma}A;\xi0\rangle =-\langle T_m|H_t|AD_{\sigma}A;\xi0\rangle$ hold, 
which imply that in addition $\mathcal{F}_{mg}^{O K}\left(q=0\right)=0$. 
Conversely, this implies that magnetic scattering at the O $K$-edge is {\it allowed} at $q=0$ and its vicinity when $\phi \neq 180^o$  
and inversion symmetry at the O-site is broken.

It is interesting to perform an analogue symmetry analysis
 for magnetic RIXS at the $L$-edge RIXS. 
The scattering amplitude $\mathcal{F}_{mg}^{L}$ is proportional to 
$c^*c_m\langle T_m | R |S\rangle$ 
and it can easily be verified that $I_{TM}$ does not give rise to 
any selection because both $|S\rangle$ and $|T_m\rangle$ have 
the same parity under $I_{TM}$. 
Interestingly, for straight bonds with $\phi=180^{\circ}$, 
where $I_{\textrm{O}}$ is a symmetry of the system, 
$\mathcal{F}_{mg}^{L}\left(q=0\right)=0$. 
This is related to the fact that for $I_{\textrm{O}}$ 
the singlet and triplet states have different parities. 
These selection rules are of course directly related to the experimental and 
theoretical observations that on iridates and cupartes 
the magnetic $L$-edge RIXS intensity peaks at the edge of the BZ
\cite{Kim2012a,Kim2012b,Gretarsson2013,Braicovich2009a,Braicovich2009b}.

{\it Exact diagonalization results ---}
In order to test these symmetry-based selection rules and 
to determine in addition their dependence 
in the incoming and outgoing polarization of x-ray photons, 
we perform an exact diagonalization study of the relevant microscopic model Hamiltonian 
for iridates and cuprates on small clusters. 
The generic Hamiltonian is $ H=\sum_{ij} H^{\textrm{TM}}_i +H^{\textrm{O}}_j+H^t_{ij},$ 
where $H^{\textrm{TM}}_i$ and $H^{\textrm{O}}_j$ refer to 
the local Hamiltonian of TM and O ions on neighboring sites 
$i$ and $j$ and $H^t_{ij}$ captures the hopping of electrons on the connecting TM-O bond
\cite{SM1}. 
Further details are provided in the supplementary material.

\begin{figure}
\centering
\includegraphics[width=\columnwidth]{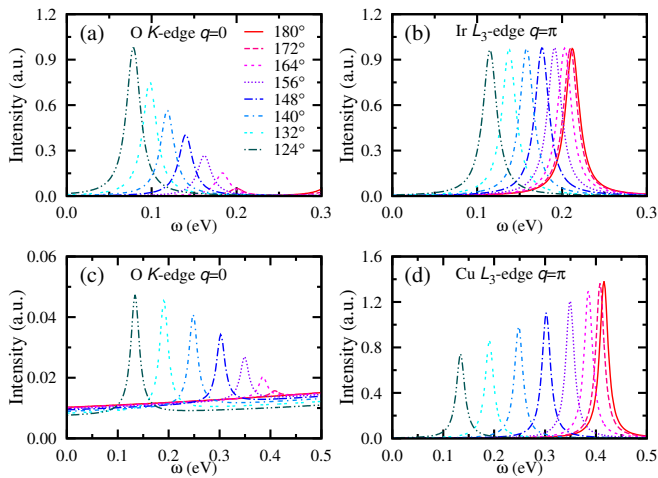}
\caption
{ (Color online) Magnetic RIXS spectra calculated by ED 
as a function of TM-O-TM bond angle $\phi$. 
(a) Iridate O $K$-edge at $q=0$, (b) Ir $L_3$-edge at $q=\pi$ , 
(c) Cuprate O $K$-edge RIXS at $q=0$, (d) Cu $L_3$-edge at $q=\pi$. 
The polarizations are $\epsilon=x$ and $\epsilon^{\prime}=z$.
Note that as a function of $\phi$ both the effective magnetic interactions and the RIXS intensity change.
}
\label{fig2}
\end{figure}

The RIXS spectra, calculated as a function of the TM-O bond angle $\phi$ 
are shown in Fig.~\ref{fig2}, where the two top panels are the results 
for an Ir$^{4+}$ cluster and the two lower panels are for a Cu$^{2+}$ cluster, 
with on the left the magnetic scattering at the oxygen $K$-edges and 
on the right RIXS intensities at the TM $L$-edges~\cite{SM2}. 
There are many charge excitations at higher energy, but we focus on only the low-energy, 
magnetic part of the excitation spectra. The overall trends are very similar. 
First of all, it is clear that when $\phi$ starts to deviate from 180$^{\circ}$, 
the energy of the magnetic excitations goes down, 
which is due to the reduction of the effective exchange interaction 
between the (iso)spins, as the hopping integrals directly depend on the bond-angle. 
This is equally true for RIXS at the Ir or Cu $L$-edges and at the corresponding O $K$-edges.
A big difference is, however, that for straight bonds, where $\phi=180^{\circ}$, 
the O $K$-edge magnetic scattering intensity vanishes, 
whereas the TM $L$-edge intensity is finite in both cases. 
This is the result of the selection rules derived above. 
Only when the bond-angle $\phi$ begins to deviate from 180$^{\circ}$, 
magnetic RIXS intensity starts to build up at the O $K$-edge. 
It should be noted, however, that for typical parametrizations of quasi-2D 
perovskite iridates and cuprates, 
the magnetic scattering intensity at the oxygen edge is at least $\sim 20$ times larger 
in the irradiate compared to the cuprate. 
This reflects the fact that the spin-orbit coupling in iridates is much larger.

\begin{figure}[t]
\centering
\includegraphics[width=\columnwidth]{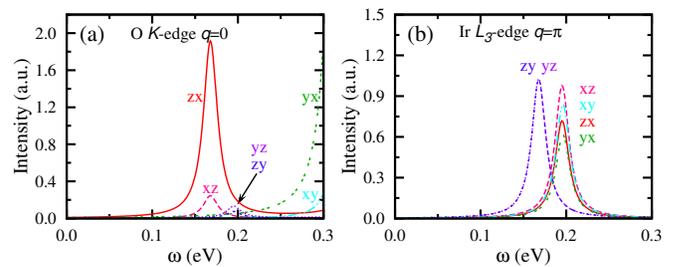}
\caption
{ (Color online)  Polarization dependence of magnetic RIXS excitations 
for an iridate cluster with an Ir-O-Ir bond-angle of $\phi=158^{\circ}$. 
(a) O $K$-edge at $q=0$, (b) Ir $L_3$-edge at $q=\pi$. The polarization condition $zx$ labels $\epsilon =z$, $\epsilon^\prime =x$.
}
\label{fig3}
\end{figure}

It is interesting to analyze the polarization dependence of the RIXS amplitude 
for an iridate with buckled bonds. 
The polarization direction of the incoming ($\epsilon$) and outgoing 
($\epsilon^\prime$) x-ray are defined in the coordinate system of Fig.~\ref{fig1} 
and the resulting RIXS spectra shown in Fig.~\ref{fig3}. 
It should first of all be noted that there are two sets of magnetic excitations 
because of the additional anisotropic interactions
(Dzyaloshinskii-Moriya and Kitaev exchange terms) that are generated 
by the buckling of bonds~\cite{Jackeli2009,BHKim2012}, 
splitting up the excited magnetic states. 
At the Ir $L$-edge, one set of magnetic modes is picked up in $yz$ and $zy$ polarization, 
and the other four orthogonal polarization conditions pick up the other magnetic mode. The magnetic scattering is only present in cross-polarization conditions,
which reflects the fact that in SM scattering angular momentum is transferred from the x-ray photon to the magnon.
This is also reflected in the O $K$-edge, RIXS, 
where depending on the specific polarization conditions, 
either of the magnetic modes is picked up ($zx$, $xz$ versus $yz$, $zy$), 
or none ($yx$ and $xy$). Note that the magnetic scattering intensity is by far largest for the situation of incoming polarization in the $z$-direction and outgoing polarization along the $x$-axis.
Note that in the spectra we do not show the elastic contribution to 
the resonant scattering, which in principle produces a zero-loss peak. 

\begin{figure}[t]
\centering
\includegraphics[width=.7\columnwidth]{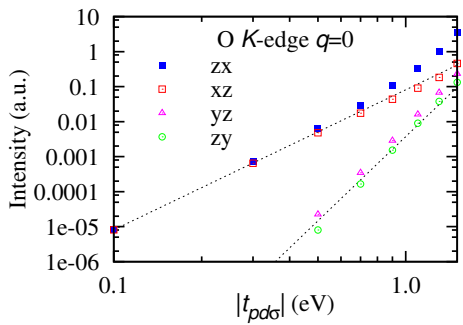}
\caption
{ (Color online)
   Magnetic RIXS intensity at the O $K$-edge as a function of the Ir-O hopping 
$t_{pd\sigma}$ on a log-log scale, for fixed $\phi=158^{\circ}$.
   The two dotted lines represent proportionalities to $t_{pd\sigma}^4$ and $t_{pd\sigma}^8$.
}
\label{fig4}
\end{figure}

It is clear that the calculated magnetic RIXS intensity at the oxygen $K$-edge of iridates, 
apart from symmetry, critically depends on an interplay of the spin-orbit coupling 
and the hybridization between oxygen $p$ and iridium $d$ states 
-- without either one or the other the magnetic RIXS intensity vanishes. 
In the cluster the Ir-O hopping is parametrized by $t_{pd\sigma}$ and 
the magnetic scattering intensity as a function of  $t_{pd\sigma}$ 
is shown in Fig.~\ref{fig4} on a log-log scale for a fixed bond-angle of $158^{\circ}$. 
It makes clear that there are two types of contributions to the intensity, 
scaling with $t_{pd\sigma}^4$ and $t_{pd\sigma}^8$, respectively, in the weak hopping limit. Moreover, which contribution is picked up strongly depends on polarization conditions of the experiment. 
This is of course very different from magnetic RIXS intensities 
at transition metal $L$-edges and their polarization dependence, 
which are in leading order finite and independent of $t_{pd\sigma}$.

{\it Summary and conclusions ---}
We have shown on the basis of a symmetry analysis and exact diagonalization study, 
that magnon scattering at O $K$-edges becomes allowed 
when the inversion symmetry at the O-site is broken, for instance by a buckling of bonds. 
For cuprates the resulting magnetic RIXS intensity is tiny 
but iridates may have the spin-orbit energy scale that is needed and 
also can fulfill the symmetry requirements. 
The first type of material that may come to mind to observe single-magnon modes 
with oxygen $K$-edge RIXS, are the quasi-2D iridates as Sr$_2$IrO$_4$ and Ba$_2$IrO$_4$, 
in which $L$-edge RIXS has already established the presence of distinct magnon modes
\cite{Kim2012a}.
One can easily extent our symmetric analysis to the corresponding 2D lattices, 
which are still invariant under $I_{TM}$ and break $I_{\textrm{O}}$ when the IrO$_4$ octahedra buckle. 
But even if the spin-orbit coupling and the Ir-O hybridization $t_{pd\sigma}$ 
are significant in these materials, the Ir-O-Ir buckling angle for these systems 
is rather small, or even vanishing~\cite{Sala2014}.  
A better candidate system is the post-perovskite CaIrO$_3$ 
in which a strong antiferromagnetic interaction along the $c$ axis ($J_c$) is present, 
where at the same time the corner-sharing octahedra are significantly buckled. 
Along the $a$ axis the octahedra are edge-sharing, 
which leads to weak ferromagnetic interactions so that magnetically CaIrO$_3$ 
is a quasi-one-dimensional system~\cite{Bogdanov2012,Ohgushi2013}, with $|J_c/J_a|>>1$ 
-- it thereby becomes an ideal candidate to observe single-magnon RIXS at the oxygen $K$-edge.

{\it Acknowledgements ---}
We thank Thorsten Schmitt and Giniyat Khaliullin for fruitful discussions. B.H.K was supported by the Basic Science Research Program through the National Research Foundation of Korea (NRF) funded by the Ministry of Education, Science and Technology (2013R1A6A3A0302053).

\renewcommand{\thefigure}{S\arabic{figure}}
\renewcommand{\thetable}{S\arabic{table}}

\renewcommand{\thefigure}{S\arabic{figure}}
\renewcommand{\thetable}{S\arabic{table}}

\setcounter{figure}{0}
\setcounter{table}{0}

\onecolumngrid

\newpage

\begin{center}
{\bf \Large 
{\it Supplemental Material:}\\
Resonant inelastic x-ray scattering on single-magnons at oxygen $K$-edges
}
\end{center}

\section{Microscopic Hamiltonian and numerical method}

To investigate the magnetic response of the oxygen (O) $K$- and 
transition-metal (TM) $L$-edge resonant inelastic x-ray scattering (RIXS) 
spectra of iridates and cuprates,
we considered the periodic TM1-O1-TM2-O2 cluster (see Fig.~1) 
which is described by following Hamiltonian:
\begin{equation}
\label{H_eq}
H=\sum_{ij}H_i^{\textrm{TM}}+H_j^{\textrm{O}}+H_{ij}^t,
\end{equation}
where $H_i^{\textrm{TM}}$ and $H_j^{\textrm{O}}$ refer to local energy of 
$i$-th TM and $j$-th O ions, respectively, and 
$H_{ij}^t$ represents the hopping Hamiltonian between neighboring TM-O bonds.
The local Hamiltonians of TM and O sites are as follows.
\begin{align}
\label{eq1}
H^{\textrm{TM}} &=\sum_{\mu\sigma} \epsilon_{\mu} n_{\mu\sigma}+
   \lambda \sum_{\mu\nu\sigma\sigma^{\prime}}
   (\mathbf{l}\cdot\mathbf{s})_{\mu\sigma,\nu\sigma^{\prime}} 
   c_{\mu\sigma}^{\dagger}c_{\nu\sigma^{\prime}} \nonumber 
  + \frac{1}{2}\sum_{\sigma\sigma^{\prime}\mu\nu} 
  U_{\mu\nu} c_{\mu\sigma}^{\dagger}c_{\nu\sigma^{\prime}}^{\dagger}
  c_{\nu\sigma^{\prime}}c_{\mu\sigma}  
  + \frac{1}{2}\sum_{\substack{\sigma \sigma^{\prime} \\ \mu\ne\nu}}
  J_{\mu\nu}c_{\mu\sigma}^{\dagger}c_{\nu\sigma^{\prime}}^{\dagger}
  c_{\mu\sigma^{\prime}}c_{\nu\sigma}  \nonumber \\
  +&\frac{1}{2}\sum_{\substack{\sigma \\ \mu\ne\nu}} 
  J_{\mu\nu}^{\prime} c_{\mu\sigma}^{\dagger}c_{\mu\bar{\sigma}}^{\dagger}
  c_{\nu\bar{\sigma}}c_{\nu\sigma},  
\end{align}
where $U_{\mu\mu}=U$, $U_{\mu\ne\nu}=U-2J_H$, and 
$J_{\mu\nu}=J_{\mu\nu}^{\prime}=J_H$.
$\bar{\sigma}$ stands for the opposite sign of $\sigma$.
We set $\epsilon_{\mu}$ so that $-1/3\Delta_t$ for $yz$ and $zx$,
$2/3\Delta_t$ for $xy$, $10Dq-3/2\Delta_t$ for $z^2$, and 
$10Dq+3/2\Delta_t$ for $x^2-y^2$.
\begin{align}
H^{\textrm{O}} &=\epsilon_{p} \sum_{m\sigma} n_{m\sigma}+
  \frac{1}{2}\sum_{\sigma\sigma^{\prime}mn} U_{mn} 
  c_{m\sigma}^{\dagger}c_{n\sigma^{\prime}}^{\dagger}
  c_{n\sigma^{\prime}}c_{m\sigma}  
  + \frac{1}{2}\sum_{\substack{\sigma \sigma^{\prime} \\ m\ne n}}
  J_{mn}c_{m\sigma}^{\dagger}c_{n\sigma^{\prime}}^{\dagger}
  c_{m\sigma^{\prime}}c_{n\sigma} 
  + \frac{1}{2}\sum_{\substack{\sigma \\ m \ne n}} 
  J_{mn}^{\prime} c_{m\sigma}^{\dagger}c_{m\bar{\sigma}}^{\dagger}
  c_{n\bar{\sigma}}c_{n\sigma},  
\end{align}
where $U_{mm}=U_p$, $U_{m\ne n}=U_p-2J_p$, and 
$J_{mn}=J_{mn}^{\prime}=J_p$.
$\epsilon_{p}$ is determined for the charge transfer energy ($\Delta$)
to be the minimum energy between $d^{5}p^6$ ($d^{9}p^6$) and $d^{6}p^5$ ($d^{10}p^5$)
configurations for iridates (cuprates).
Hopping Hamiltonian is given by 
\begin{equation}
H_{ij}^t=\sum_{\mu m\sigma} \big( t_{\mu m}^{ij} c^{\dagger}_{jm\sigma}
   c_{i\mu\sigma} + h.c.\big),
\end{equation}
where $t_{\mu m}^{ij}$ is a function of two parameters $t_{pd\sigma}$ and
$t_{pd\pi}$ based on the Slater-Koster theory~\cite{Slater}.

We restricted the Hilbert space in order to include all possible states whose
configuration is $d^n$-$d^n$-$p^6$-$p^6$, $d^{n-1}$-$d^{n+1}$-$p^6$-$p^6$, 
$d^n$-$d^{n+1}$-$p^5$-$p^6$, $d^{n+1}$-$d^{n+1}$-$p^5$-$p^5$, or 
$d^{n+1}$-$d^{n+1}$-$p^4$-$p^6$ where $n=5$ ($n=9$) for iridate (cuprate).
We took into account the following relevant local multiplets 
to describe electronic states of Ir (Cu) ions: 
6 (4) lowest multiplets for $d^5$ ($d^9$), 14 (6) lowest multiplets for $d^{4}$ ($d^8$), 
and 1 (1) lowest mutliplets for $d^6$ ($d^{10}$).
Because only $t_{2g}$-hole ($e_g$-hole) states play a dominant role in 
low energy excitations of iridates (cuprates), 
they are enough to describe the single-magnon contribution in the RIXS spectra
\cite{BHKim1,BHKim2}.
We constructed the Hilbert space so that each state can be expressed as 
the direct product of local multiplets of TM and O ions.

\begin{table}[b]
\caption
{Physical parameters for iridates in units of eV, consistent with previous work~\cite{BHKim1}.
}
\label{ir_para}
\begin{ruledtabular}
\begin{tabular}{c c c c c c c c c c}
 $10Dq$ & $\Delta_{t}$ & $U$ & $J_H$ & $\lambda$ & 
 $\Delta$ & $U_p$ & $J_p$ & $t_{pd\sigma}$ & $t_{pd\pi}$ \\
\hline
3.0 & 0.10 & 1.86 & 0.50 & 0.4 & 3.3 & 0.5 & 0.1 & $-1.5$ & $-0.46t_{dp\sigma}$ \\
\end{tabular}
\end{ruledtabular}
\end{table}

\begin{table}[h]
\caption
{Physical parameters for cuprates in units of eV, consistent with previous work~\cite{Imada}.
}
\label{cu_para}
\begin{ruledtabular}
\begin{tabular}{c c c c c c c c c c}
 $10Dq$ & $\Delta_{t}$ & $U$ & $J_H$ & $\lambda$ & 
 $\Delta$ & $U_p$ & $J_p$ & $t_{pd\sigma}$ & $t_{pd\pi}$ \\
\hline
1.5 & $-0.10$ & 8.0 & 0.5 & 0.1 & 2.5 & 0.5 & 0.1 & $-1.0$ & $-0.46t_{dp\sigma}$ \\
\end{tabular}
\end{ruledtabular}
\end{table}

We adopted physical parameters as shown in table~\ref{ir_para} (table~\ref{cu_para}) 
for iridates (cuprates) and solved the Hamiltonian by using the exact diagonalization. 
We calculated the RIXS spectra as following Kubo formula:
\begin{equation}
\label{RX_eq}
I = \frac{1}{\Lambda^2} \sum_n |\langle n|R|g\rangle|^2
      \delta(\omega-E_n+E_g)
\end{equation}
where $E_g$ ($E_n$) is the energy of the ground state $|g\rangle$ 
(the excited state $|n\rangle$) and $R$ is the RIXS scattering operator.

\section{RIXS spectra for iridium $L$-edge RIXS at $q=0$}

\begin{figure}[ht]
\centering
\includegraphics[width=10.0 cm]{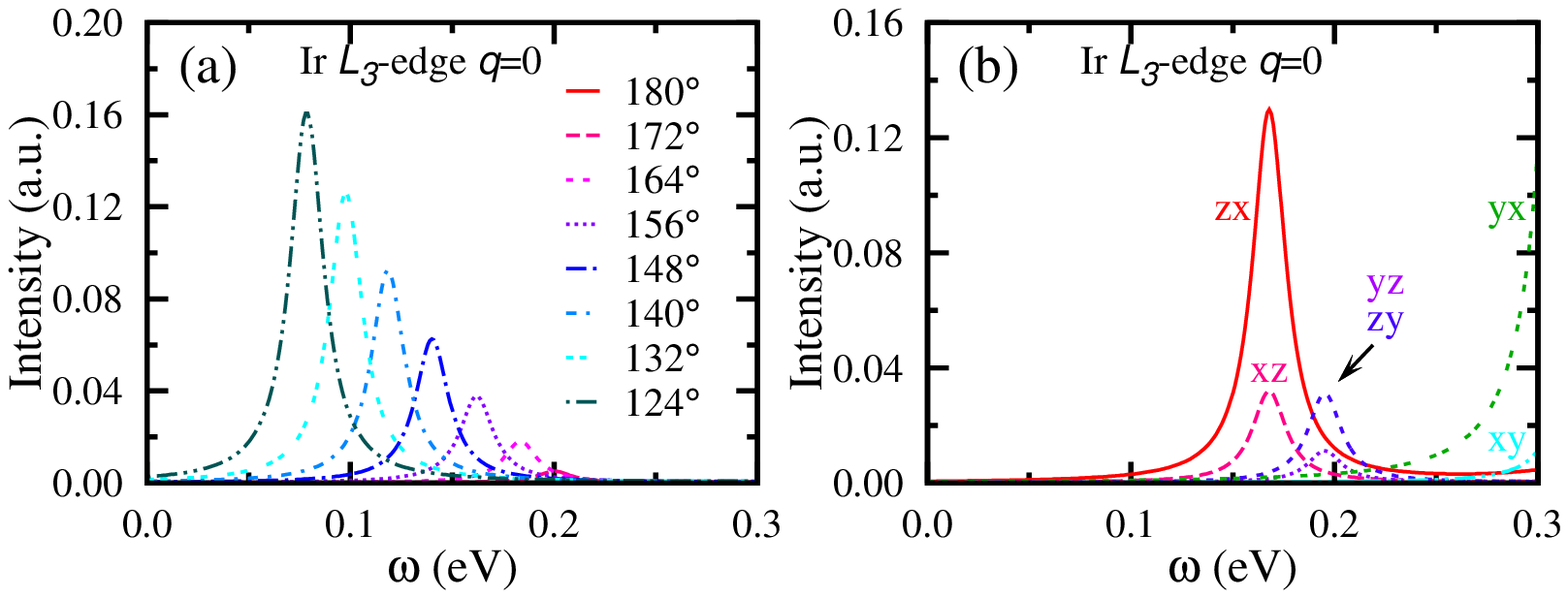}
\caption
{ (Color online)
   Magnetic RIXS spectra of the Ir $L_3$-edge at $q=0$ as a function of 
   (a) Ir-O-Ir bond angle and (b) x-ray polarization.
   $\phi=158^{\circ}$ is set in (b).
}
\label{sfig1}
\end{figure}

As shown in table~II, the RIXS scattering operator of the Ir $L$-edge at $q=0$ has
same symmetric property as that of the O $K$-edge. It can be also expected for
the Ir $L$-edge RIXS to give rise to the magnetic peak at $q=0$.
Accordingly our numerical calculation yields nonzero magnetic RIXS peak of the Ir $L$-edge at $q=0$.
Figure~\ref{sfig1} presents the magnetic RIXS peak of the $L_3$-edge at $q=0$
with respect to bond angle and x-ray polarization.
Overall features are totally same as those in the O $K$-edge (see Fig.~2(a) and 3(a)).
The magnetic peak at $q=0$ is only allowed when the Ir-O-Ir bond is buckled.
The magnetic modes is dominantly picked up at $zx$ and $xz$ polarizations.
Neither $yx$, nor $xy$, nor $xx$ polarization gives any magnetic peak.
Thus the broken inversion with oxygen center can induce
the magnetic excitation at center of the BZ even though its intensity is
much more suppressed than that at the edge.

\begin{figure}[ht]
\centering
\includegraphics[width=6.0 cm]{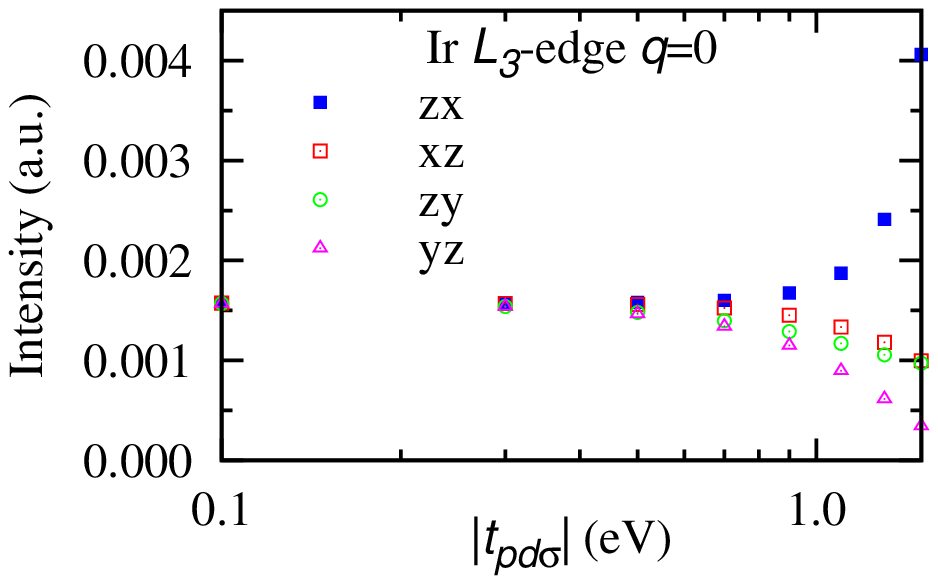}
\caption
{ (Color online)
   Magnetic RIXS intensity of the Ir $L_3$-edge at $q=0$ 
   as a function of the Ir-O hopping for fixed $\phi=158^{\circ}$.
}
\label{sfig2}
\end{figure}

The $L$-edge RIXS process, however, is not directly related to O states.
For this reason, the magnetic RIXS intensity at the Ir $L$-edge is almost
independent with the Ir-O hopping strength in contrast with the O $K$-edge, 
of which intensity is proportional to $t_{pd\sigma}^4$ or $t_{pd\sigma}^8$.
Figure~\ref{sfig2} describes this feature well. In weak hopping limit,
total magnetic RIXS intensity is almost constant.
In strong hopping limit, of course, even Ir $L$-edge is remarkably affected 
due to considerable overlap between local Ir multiplet states and charge-transfer states.

\end{document}